\documentclass{article}
\usepackage{amsmath}[1996/11/01]
\usepackage{amssymb,amsthm,amsxtra}

\def\[#1\]{\begin{equation}#1\end{equation}}
\makeatletter
\def\beq{%
   \relax\ifmmode
      \@badmath
   \else
      \ifvmode
         \nointerlineskip
         \makebox[.6\linewidth]%
      \fi
      $$
   \fi
}
\def\eeq{%
   \relax\ifmmode
      \ifinner
         \@badmath
      \else
         $$
      \fi
   \else
      \@badmath
   \fi
   \ignorespaces
}

\def\enddisplaymath{\eeq\global\@ignoretrue}
\makeatother

\newtheorem{thm}{Theorem}
\newtheorem{cor}[thm]{Corollary}
\newtheorem{lem}[thm]{Lemma}

\theoremstyle{remark}
\newtheorem*{rem}{Remark}
\newtheorem*{eg}{Example}
\newtheorem{rems}{Remark}[thm]

\theoremstyle{definition}
\newtheorem{defn}{Definition}

\numberwithin{equation}{section}
\numberwithin{thm}{section}

\DeclareMathOperator{\Tr}{Tr}
\DeclareMathOperator{\GL}{GL}
\DeclareMathOperator{\Exp}{{\bf E}}

\DeclareMathOperator{\Hom}{Hom}
\DeclareMathOperator{\Mat}{Mat}

\DeclareMathOperator{\wgt}{wt}

\newcommand{\C}{\mathbb C}

\newcommand{\Z}{\mathbb Z}

\begin{document}

\title{A semidefinite program for distillable entanglement}
\author{Eric M. Rains\\AT\&T Research
                     \\rains@research.att.com}
\date{April 12, 2001}
\maketitle

\begin{abstract}
We show that the maximum fidelity obtained by a p.p.t. distillation
protocol is given by the solution to a certain semidefinite program.  This
gives a number of new lower and upper bounds on p.p.t. distillable
entanglement (and thus new upper bounds on 2-locally distillable
entanglement).  In the presence of symmetry, the semidefinite program
simplifies considerably, becoming a linear program in the case of isotropic
and Werner states.  Using these techniques, we determine the
p.p.t. distillable entanglement of asymmetric Werner states and ``maximally
correlated'' states.  We conclude with a discussion of possible
applications of semidefinite programming to quantum codes and 1-local
distillation.

\end{abstract}

\section{Introduction}

One of the central problems of quantum information theory is entanglement
distillation (\cite{Bennettetal}, \cite{Rains:T}): the production
of (approximate) maximally entangled states from a collection of
non-maximally entangled states.  Of particular interest are 1-locally
distillable entanglement and 2-locally distillable entanglement (the
amount of entanglement that can be distilled using local operations
and a 1-way (2-way) classical channel).  Nearly all of the known upper
bounds on 1- or 2-locally distillable actually apply to a larger class
of operations, known as p.p.t. (positive partial transpose) operations
\cite{Rains:T}.  This motivates our present study of p.p.t. distillable
entanglement.

We study distillable entanglement via a more refined quantity, the
``fidelity of distillation'', which measures how close one can come
to producing a $K$-dimensional maximally entangled state from a given
input.  In Theorem \ref{thm:sdp1} below, we show that the fidelity
of p.p.t. distillation can be expressed as the solution to a certain
semidefinite program (see \cite{SDP} for a survey of semidefinite
programming).  Then any feasible solution to the dual problem (Theorem
\ref{thm:sdp2}) gives us an upper bound on fidelity of distillation.

The rest of the paper is devoted to an exploration of the consequences of
this semidefinite program.  Section 4 gives a number of results that hold
in general, including a new bound combining the bounds of \cite{Rains:U}
and \cite{Horodecki1999}, and a theorem to the effect that maximally
entangled states cannot be used to catalyze fidelity of
p.p.t. distillation.  In section 5, we show that the semidefinite program
simplifies in the presence of symmetries; in some cases (e.g., isotropic
states, Werner states), this simplification turns the semidefinite program
into a {\it linear} program.  In the case of asymmetric Werner states, this
linear program can be solved exactly, showing that the upper bound of
\cite{Horodecki1999} is tight in that case.  Section 6 sketches a technique
for producing asymptotic lower bounds, which we then use to strengthen the
hashing lower bound \cite{Bennettetal} in the p.p.t. case.  We also use
this technique to partially resolve a conjecture of \cite{Rains:U} by
determining the p.p.t. distillable entanglement of ``maximally correlated''
states.  Finally, in section 7, we consider possible applications of
semidefinite programming to the problems of quantum codes and 1-local
distillation.  In particular, using the techniques of section 5, we
give a new derivation of the linear programming bound for quantum codes
\cite{ShorLaflamme}, \cite{Rains:K}, \cite{Rains:W}.

\section{Operators, superoperators and operations}

If $V$ is a Hilbert space, we denote by ${\cal H}(V)$ the space of
Hermitian operators on $V$.  We also let ${\cal P}(V)\subset {\cal H}(V)$
denote the convex cone of positive semi-definite Hermitian operators; we
will freely write $A\ge B$ to mean $A-B\in {\cal P}(V)$.  A state is then
an element of ${\cal P}(V)$ of trace 1.  Quantum information theory can be
thought of as studying the behavior of these concepts under tensor
products.

Given an operator $A\in {\cal H}(V\otimes W)$, we define the ``partial
trace'' $\Tr_V(A)$ to be the (unique) operator in ${\cal H}(W)$ such
that
\[
\Tr(\Tr_V(A) B) = \Tr(A (B\otimes 1)),
\]
for all $B\in {\cal H}(W)$.  Similarly, given a choice of basis for $W$,
we can define the partial transpose $A^{\Gamma_W}$ by
\[
\Tr(A^{\Gamma_W} (B\otimes C)) = \Tr(A (B\otimes C^t)),
\]
where $B\in {\cal H}(V)$, $C\in {\cal H}(W)$, and $C^t$ is the
transpose of $C$ with respect to the chosen basis.  Both of these
transformations extend by linearity to non-Hermitian operators as well.

A positive operator $C\in {\cal P}(V\otimes W)$ is said to be ``separable''
if it can be written in the form
\[
C = \sum_i A_i\otimes B_i,
\]
with $A_i\in {\cal P}(V)$, $B_i\in {\cal P}(W)$; in other words,
\[
C\in {\cal P}(V)\otimes {\cal P}(W).
\]
Similarly, $C$ is said to be p.p.t. (positive partial transpose) if
\[
C \in {\cal P}(V\otimes W) \cap {\cal P}(V\otimes W)^{\Gamma_W};
\]
note that this does not depend on the choice of basis in $W$.
We also recall that every p.p.t. operator is separable:
\[
{\cal P}(V)\otimes {\cal P}(W) \subset {\cal P}(V\otimes W) \cap {\cal
P}(V\otimes W)^{\Gamma_W}.
\]

A ``superoperator'' from $V$ to $V'$ is a linear transformation from ${\cal
H}(V)$ to ${\cal H}(V')$.  The space of superoperators can be naturally
identified with ${\cal H}(V\otimes V')$; to a superoperator $\Psi$
corresponds the unique operator $\Omega(\Psi)$ such that
\[
\Tr(B \Psi(A)) = \Tr\left(\Omega(\Psi) \left(A\otimes B\right)\right).
\]
We also define the adjoint superoperator $\Psi^*$ by
\[
\Tr(A \Psi^*(B)) = \Tr(B\Psi(A)).
\]
Note that
\begin{align}
\Psi(A) &= \Tr_V(\Omega(\Psi) (A\otimes 1))\\
\Psi^*(B) &= \Tr_{V'}(\Omega(\Psi) (1\otimes B))
\end{align}
and, if $\Psi_1:{\cal H}(V)\to {\cal H}(V')$ and $\Psi_2:{\cal H}(V')\to
{\cal H}(V'')$, then
\[
\Omega(\Psi_2\circ \Psi_1)
=
(\Psi_1^*\otimes 1)(\Omega(\Psi_2))
=
(1\otimes \Psi_2)(\Omega(\Psi_1))
=
\Tr_{V'} ((\Omega(\Psi_1) \otimes 1_{V''})(1_V\otimes \Omega(\Psi_2))).
\]
Of particular interest is the (self-adjoint) superoperator $A\mapsto A^t$;
in that case, we find
\[
\Omega(t) = \sum_{i,j} (v_i\otimes v_i) (v_j\otimes v_j)^\dagger \ge 0.
\]

A superoperator is said to be ``positive'' if $\Psi(A)\ge 0$ whenever $A\ge
0$, and ``trace-preserving'' if $\Psi^*(1)=1$; equivalently,
$\Tr_{V'}(\Omega(\Psi))=1$.  A superoperator is ``completely positive'' if it
satisfies any of the following equivalent conditions:
\begin{itemize}
\item{(1)} $1_V\otimes \Psi$ is positive
\item{(2)} For all Hilbert spaces $W$, $1_W\otimes \Psi$ is positive
\item{(3)} There exist operators $A_i\in \Hom(V,V')$ such that
\[
\Psi(A) = \sum_i A_i A A_i^\dagger.
\]
\item{(4)} For any (some) basis of $V$, the partial transpose
$\Omega(\Psi)^{\Gamma_V}$ is positive semi-definite.
\end{itemize}
Clearly $2\implies 1$, and $3\implies 2$ is straightforward.  To see
$1\implies 4$, it suffices to observe that
\[
\Omega(\Psi)^{\Gamma_V} = (1_V\otimes \Psi)(\Omega(t)) \ge 0.
\]
Finally, $4\implies 3$ follows by taking an eigenvalue decomposition of
$\Omega(\Psi)^{\Gamma_V}$.  Since the operators we will be dealing with in
the sequel are mostly completely positive, we define
$\Omega'(\Psi)=\Omega(\Psi)^{\Gamma_V}$, and use this to identify the space
of superoperators with ${\cal H}(V\otimes V')$.  Thus the set of completely
positive superoperators is identified with ${\cal P}(V\otimes V')$.  An
``operation'' is defined to be a completely positive, trace-preserving
superoperator; we denote the (convex) set of operations from $V$ to $V'$ by
$Op(V,V')$.\footnote{This differs somewhat from the definition of operation
given in \cite{Rains:T}, in that we are assuming operations to be
``non-measuring'', but by the main result of that paper, this incurs no
loss of generality when studying entanglement distillation.}

On tensor product spaces, there are several classes of operations of
interest, which can be defined in terms of the convex sets ${\cal P}$ and
$Op$ as follows:
\begin{itemize}
\item{} $\epsilon$-local: ${\cal C}_\epsilon=Op(V,V')\otimes Op(W,W')$.
\item{} $1$-local: ${\cal C}_1=({\cal P}(V\otimes V')\otimes Op(W,W'))\cap
Op(V\otimes W,V'\otimes W')$.
\item{} $1'$-local: ${\cal C}_{1'}=(Op(V,V')\otimes {\cal P}(W\otimes W'))\cap
Op(V\otimes W,V'\otimes W')$.
\item{} separable: ${\cal C}_{\$}=({\cal P}(V\otimes V')\otimes {\cal
P}(W\otimes W'))\cap Op(V\otimes W,V'\otimes W')$.
\item{} p.p.t.: ${\cal C}_\Gamma=Op(V\otimes W,V'\otimes W')\cap
Op(V\otimes W,V'\otimes W')^{\Gamma_{W\otimes W'}}$.
\end{itemize}
We also have the class of 2-local operations, defined by allowing arbitrary
compositions of 1-local and $1'$-local operations.  For a different
approach to defining these classes, see \cite{Rains:T}.  We recall
\[
{\cal C}_\epsilon \subset {\cal C}_1,{\cal C}_{1'}\subset {\cal C}_2\subset
{\cal C}_{\$}\subset {\cal C}_\Gamma,
\]
with all inclusions strict in general.  (The class ${\cal C}_\epsilon$,
not discussed in \cite{Rains:T}, is simply the closure of the class of
local operations under convex linear combinations (i.e., shared randomness).)

From a physical perspective, the only natural classes are those of
$(\epsilon,1,1',2)$-local operations.  The difficulty, however, is that in
none of these cases do we have an effective way to decide whether a given
operation belongs to the class; this is especially true in the case of
2-local operations.  Thus the class of separable operations is important as
a simplification of the class of 2-local operations, while the class of
p.p.t. operations is important as the smallest class containing the 2-local
class for which we can effectively decide membership.  For instance, all of
the known upper bounds on 2-locally distillable entanglement are really
bounds on p.p.t. distillable entanglement; to a large extent this even
applies to upper bounds on 1-locally distillable entanglement.
Similarly, a lower bound on p.p.t. distillable entanglement provides
a limit on how far the current methods can take us.

\section{Fidelity of distillation}

For any integer $K>0$, we define the ``maximally entangled'' state
$\Phi(K)\in {\cal H}(\C^{K}\otimes \C^{K})$ by
\[
\Phi(K) = \frac{1}{K} \Omega'(1_{\C^K})
=
\frac{1}{K} \sum_{1\le i,j\le K} (e_i\otimes e_i)(e_j\otimes e_j)^\dagger.
\]
Given any other state $\rho$, the ``fidelity'' of $\rho$ is
defined by
\[
F(\rho) = \Tr(\Phi(K) \rho).
\]

\begin{defn}\label{def:FGamma1}
Let $\rho\in {\cal P}(V\otimes W)$ be a state, and let $K>0$ be an integer.
The ``fidelity of $K$-state p.p.t. distillation'' $F_\Gamma(\rho;K)$
is defined by
\[
F_\Gamma(\rho;K)
=
\max_\Psi F(\Psi(\rho)),
\]
where $\Psi$ ranges over all p.p.t. operations from ${\cal H}(V\otimes W)$
to ${\cal H}(\C^K\otimes \C^K)$.
\end{defn}

\begin{rem}
We can define $F_\epsilon$, $F_1$, $F_2$, etc., similarly.
\end{rem}

This is a refinement of the concept of distillable entanglement;
indeed, we can define (see \cite{Bennettetal}, \cite{Rains:T}):

\begin{defn}
Let $\rho$ be as above.  The p.p.t. distillable entanglement
$D_\Gamma(\rho)$ of $\rho$ is defined to be the supremum of
all positive numbers $r$ such that
$\lim_{n\to\infty} F_\Gamma(\rho^{\otimes n};\lfloor 2^{rn}\rfloor) = 1$.
\end{defn}

Thus a study of $F_\Gamma$ is likely to provide insights into $D_\Gamma$,
as we shall indeed find below.

We first observe that the optimization problem defining $F_\Gamma$
can be rewritten as an optimation over operators:

\begin{thm}\label{thm:sdp1}
For any state $\rho$ and any positive integer $K$,
\[
F_\Gamma(\rho;K) = \max_F \Tr(F\rho),\label{sdp:primal}
\]
where $F$ ranges over Hermitian operators such that
\[
0\le F\le 1,\ -1/K\le F^\Gamma \le 1/K.\label{sdp:primalineqs}
\]
\end{thm}

\begin{proof}
Let $\Psi$ be the operation maximizing $F(\Psi(\rho))$ in the definition of
$F_\Gamma(\rho;K)$.  Clearly, if we compose $\Psi$ with any operator of the
form $U\otimes \overline{U}$, this leaves $F(\Psi(\rho))$ unchanged.  The
same must then be true after averaging over $U(K)$ (``twirling''
\cite{Bennettetal}).  We may thus assume $\Psi = {\bf T}\circ \Psi$,
where ${\bf T}$ is the twirling superoperator.  We find
\[
{\bf T}(A) = \Tr(A\Phi(K))\Phi(K) +
\frac{1}{K^2-1}\Tr(A(1-\Phi(K)))(1-\Phi(K));
\]
${\bf T}(A)$ must have the form $a\Phi(K)+b(1-\Phi(K))$, and since
\[
\Tr({\bf T}(A))=\Tr(A)\quad\text{and}\quad \Tr({\bf
T}(A)\Phi(K))=\Tr(A\Phi(K)),
\]
we can solve for $a$ and $b$.  It follows that
\[
\Omega({\bf T})= \Phi(K)\otimes \Phi(K) + \frac{1}{K^2-1} 
(1-\Phi(K))\otimes (1-\Phi(K)) = \Omega'({\bf T})
\]
But then we compute
\begin{align}
\Omega'(\Psi) =
\Omega'({\bf T}\circ \Psi)
&=
(\Psi^*\otimes 1)(\Omega'({\bf T}))\\
&=
\Psi^*(\Phi(K))\otimes \Phi(K)
+
\frac{1}{K^2-1}
\Psi^*(1-\Phi(K))\otimes (1-\Phi(K)).
\end{align}
Setting $F=\Psi^*(\Phi(K))$, we obtain:
\[
\Omega'(\Psi)
=
F\otimes \Phi(K)
+
\frac{1}{K^2-1}
(1-F)\otimes (1-\Phi(K)).
\]
This operator is positive if and only if $F\ge 0$ and $(1-F)\ge 0$.
We also find
\begin{align}
\Omega'(\Psi)^\Gamma
&=
F^\Gamma \otimes \Phi(K)^\Gamma
+
\frac{1}{K^2-1}
(1-F^\Gamma)\otimes (1-\Phi(K)^\Gamma)\\
&=
\frac{1}{K+1} (1/K+F^\Gamma)\otimes (1+K\Phi(K)^\Gamma)/2
+
\frac{1}{K-1} (1/K-F^\Gamma)\otimes (1-K\Phi(K)^\Gamma)/2.
\end{align}
Since $(1\pm K\Phi(K)^\Gamma)/2$ are orthogonal projections, we find
that $\Omega(\Psi)^\Gamma$ is positive if and only if
\[
-1/K\le F^\Gamma\le 1/K.
\]
The theorem follows by noting
\[
F(\Psi(\rho)) = \Tr(F\rho).
\]
\end{proof}

\begin{defn}
An operator that satisfies the inequalities \eqref{sdp:primalineqs}
will be said to be {\it primal feasible} for $F_\Gamma(\rho;K)$; if it
maximizes $\Tr(F\rho)$, it will be said to be {\it primal optimal}.
\end{defn}

We will use this result to define $F_\Gamma(\rho;K)$ for all positive real
values of $K$; for an interpretation, see the remark following Corollary
\ref{cor:nocatalysis} below.

\begin{thm}\label{thm:escher}
The function $F_\Gamma$ is convex in $\rho$ and concave in $1/K$; that is,
for $0\le \pi\le 1$:
\begin{align}
F_\Gamma(\pi \rho_1+(1-\pi)\rho_2;K)
&\le
\pi F_\Gamma(\rho_1;K)+(1-\pi) F_\Gamma(\rho_2;K)\\
F_\Gamma(\rho;(K_1K_2)/(\pi K_2+(1-\pi)K_1))
&\ge
\pi F_\Gamma(\rho;K_1)+(1-\pi) F_\Gamma(\rho;K_2).
\end{align}
In particular, $F_\Gamma$ is continuous in both variables.
\end{thm}

\begin{proof}
Let $F$ be primal optimal for $F_\Gamma(\pi \rho_1+(1-\pi)\rho_2;K)$.  Then
\begin{align}
F_\Gamma(\pi \rho_1+(1-\pi)\rho_2;K)
&=
\Tr(F (\pi \rho_1+(1-\pi)\rho_2))\\
&=
\pi \Tr(F \rho_1)+(1-\pi)\Tr(F \rho_2)\\
&\le
\pi F_\Gamma(\rho_1;K)+(1-\pi) F_\Gamma(\rho_2;K).
\end{align}
Similarly, let $F_1$ and $F_2$ be primal optimal for
$F_\Gamma(\rho;K_1)$
and $F_\Gamma(\rho;K_2)$ respectively.  Then $\pi F_1+(1-\pi) F_2$
is primal feasible for $F_\Gamma(\rho;(K_1K_2)/(\pi K_2+(1-\pi)K_1))$,
thus giving the second inequality.
\end{proof}

The above optimization problem is an instance of what is known as
``semi-definite programming'' (SDP) (\cite{SDP}).  That is, it involves the
optimization of a linear function subject to the constraint that certain
operators (depending linearly on the variables) must be positive
semidefinite.  This has several consequences, including the computational
one that semi-definite programs can be solved in polynomial time (typically
polynomial in the dimension, although special structure can greatly reduce
this).  Another consequence is that there is a notion of duality for SDPs.

For a Hermitian operator $A$, we define the positive part $A_+$ and
negative part $A_-$ to be the unique positive operators such that
\[
A_+-A_- = A,\ A_+ A_- = 0.
\]
We also define $|A|=A_+ + A_-$.

\begin{thm}\label{thm:sdp2}
For any state $\rho\in {\cal H}(V\otimes W)$ and any positive real number
$K$,
\[
F_\Gamma(\rho;K) = \min_{D\in {\cal H}(V\otimes W)}
\Tr (\rho-D)_+ + \frac{1}{K}\Tr |D^\Gamma|.
\]
\end{thm}

\begin{proof}
Let $F$ be an operator satisfying the constraints above.  Then
for any operators $A$, $B$, $C$, we have:
\begin{align}
\Tr(F\rho)
&=
\Tr(A)+1/K\Tr(B+C)\notag\\
&\phantom{{}={}}{}
-\Tr((-\rho+A+B^\Gamma-C^\Gamma) F)
-\Tr(A(1-F))
-\Tr(B(1/K-F^\Gamma))
-\Tr(C(1/K+F^\Gamma))
.
\end{align}
If $A\ge 0$, $B\ge 0$, $C\ge 0$, and
\[
A\ge \rho-(B-C)^\Gamma,
\]
then the last four terms are all nonnegative, and we have
\[
\Tr(F\rho)
\le
\Tr(A)+1/K\Tr(B+C),
\]
and thus
\[
F_\Gamma(\rho;K) \le \Tr(A)+1/K\Tr(B+C).
\]
In fact, by the theory of duality for SDPs, this inequality can be
made tight, to wit:
\[
F_\Gamma(\rho;K) = \min_{A,B,C} \Tr(A)+1/K\Tr(B+C),
\]
minimizing over operators satisfying the constraints.
Upon adding a variable $D$ with $D=(B-C)^\Gamma$,
the constraints become
\[
A,B,C\ge 0,\ A\ge\rho-D,\ B+C=D^\Gamma.
\]
We thus find
\[
F_\Gamma(\rho;K) = \min_D
\left(
\min_{\substack{A\ge 0\\A\ge \rho-D}} \Tr(A)
+\frac{1}{K}
\min_{\substack{B,C\ge 0\\B-C=D^\Gamma}}\Tr(B+C)
\right).
\]
But we readily see that
\begin{align}
\min_{\substack{A\ge 0\\A\ge \rho-D}} \Tr(A)
&=
\Tr(\rho-D)_+,\\
\min_{\substack{B,C\ge 0\\B-C=D^\Gamma}} \Tr(B+C)
&=
\Tr |D^\Gamma|,
\end{align}
proving the theorem.
\end{proof}

\begin{defn}
An operator $D\in {\cal H}(V\otimes W)$ such that
\[
F_\Gamma(\rho;K) = \Tr (\rho-D)_+ + \frac{1}{K}\Tr |D^\Gamma|
\]
will be said to be {\it dual optimal} for $F_\Gamma(\rho;K)$.
\end{defn}

Thus given any operator $D$, we obtain bounds on fidelity of distillation,
and conversely any such bound can in principle be shown by choosing a
suitable operator $D$.  For instance, Theorem \ref{thm:escher} could also
be proved as follows:

\begin{proof}
If $D_1$ and $D_2$ are dual optimal for $F_\Gamma(\rho_1;K)$ and
$F_\Gamma(\rho_2;K)$, then
\begin{align}
F_\Gamma(\pi\rho_1+(1-\pi)\rho_2;K)
&\le
\Tr (\pi\rho_1+(1-\pi)\rho_2-\pi D_1+(1-\pi) D_2)_+
+
\frac{1}{K}\Tr |\pi D_1^\Gamma+(1-\pi) D_2^\Gamma|\\
&
\le \pi (\Tr(\rho_1-D_1)_+ + \Tr|D_1^\Gamma|)
+
(1-\pi) (\Tr(\rho_2-D_2)_+ + \Tr|D_2^\Gamma|)\\
&
=
\pi F_\Gamma(\rho_1;K)+(1-\pi) F_\Gamma(\rho_2;K).
\end{align}

Similarly, if $D$ is dual optimal for
$F_\Gamma(\rho;(K_1K_2)/(\pi K_2+(1-\pi)K_1))$, then
\begin{align}
F_\Gamma(\rho;(K_1K_2)/(\pi K_2+(1-\pi) K_1))
&=
\Tr (\rho-D)_+ + (\pi \frac{1}{K_1}+(1-\pi) \frac{1}{K_2})\Tr |D^\Gamma|\\
&=
\pi (\Tr (\rho-D)_+ + \frac{1}{K_1}\Tr |D^\Gamma|)
+
(1-\pi) (\Tr (\rho-D)_+ + \frac{1}{K_2}\Tr |D^\Gamma|)\\
&\ge
\pi F(\rho;K_1)+(1-\pi) F(\rho;K_2).
\end{align}
\end{proof}

\section{General results}

\begin{lem}
For any integer $d\ge 1$, and any $K>0$,
$F_\Gamma(\Phi(d);K) = \min(1,d/K)$.
\end{lem}

\begin{proof}
For $K\ge d$, take $F=d/K$, $D=\Phi(d)$.
For $K\le d$, take $F=\Phi(d)$, $D=0$.
\end{proof}

\begin{thm}\label{thm:tpbounds}
For any states $\rho_1$ and $\rho_2$, and any $K,K'>0$,
\[
F_\Gamma(\rho_1;K')F_\Gamma(\rho_2;K/K')
\le
F_\Gamma(\rho_1\otimes \rho_2;K)
\le
F_\Gamma(\rho_1;K/\Tr|\rho_2^\Gamma|).
\]
\end{thm}

\begin{proof}
For the first inequality, let $F_1$ and $F_2$ be primal optimal for
$F_\Gamma(\rho_1;K')$ and $F_\Gamma(\rho_2;K/K')$; then $F_1\otimes F_2$
is primal feasible for $F_\Gamma(\rho_1\otimes \rho_2;K)$, giving the
inequality.

For the second inequality, let $D$ be dual optimal for
$F_\Gamma(\rho_1;K/\Tr|\rho_2^\Gamma|)$.
Then, taking $D'=D\otimes \rho_2$, we have
\begin{align}
F_\Gamma(\rho_1\otimes \rho_2;K)
&\le
\Tr ((\rho_1\otimes \rho_2)-(D\otimes \rho_2))_+
+
\frac{1}{K}
\Tr|(D\otimes \rho_2)^\Gamma|\\
&=
\Tr(\rho_1-D)_+ +
\frac{\Tr|\rho_2^\Gamma|}{K}
\Tr|D^\Gamma|\\
&=
F_\Gamma(\rho_1;K/\Tr|\rho_2^\Gamma|).
\end{align}
\end{proof}

In particular, if $F_\Gamma(\rho_2;\Tr|\rho_2^\Gamma|) = 1$, then equality
holds in this theorem, taking $K' = K/\Tr|\rho^\Gamma_2|$.  Since this is
true for $\Phi(d)$,

\begin{cor}\label{cor:nocatalysis}
For all integers $d$, all $K>0$, and any state $\rho$,
\[
F_\Gamma(\rho\otimes \Phi(d);dK)
=
F_\Gamma(\rho;K).
\]
\end{cor}

\begin{rem}
This gives us another way to define $F_\Gamma(\rho;K)$ for general $K>0$.
For rational $K>0$, we can define
\[
F_\Gamma(\rho;p/q) = F_\Gamma(\rho\otimes \Phi(q);p),
\]
which is well-defined by the theorem.  Since the resulting function is
nonincreasing in $K$, there is a unique way to extend it to a
left-continuous function of $K$, which must then agree with our
earlier definition.
\end{rem}

Another example is when $\rho_2$ is p.p.t.; then $\Tr|\rho_2^\Gamma|=1$.
We have:

\begin{cor}
For all $K>0$, any state $\rho$, and any p.p.t. state $\rho'$,
\[
F_\Gamma(\rho\otimes \rho';K)
=
F_\Gamma(\rho;K).
\]
\end{cor}

\begin{cor}\label{cor:negativitybound}
For any $K>0$ and any state $\rho$,
\[
\min(1,1/K) \le F_\Gamma(\rho;K) \le \min(1,\Tr|\rho^\Gamma|/K).
\]
\end{cor}

\begin{proof}
By the theorem, we have, writing $\rho = \Phi(1)\otimes \rho$:
\[
\min(1,1/K)
=
F_\Gamma(\Phi(1);K)F_\Gamma(\rho;1)
\le
F_\Gamma(\rho;K)
\le
F_\Gamma(\Phi(1);K/\Tr|\rho^\Gamma|)
=
\min(1,\Tr|\rho^\Gamma|/K).
\]
\end{proof}

Asymptotically, the theorem becomes:

\begin{cor}\label{cor:tpbounds}
For any pair of states $\rho_1$, $\rho_2$,
\[
D_\Gamma(\rho_1)+D_\Gamma(\rho_2)
\le
D_\Gamma(\rho_1\otimes \rho_2)
\le
D_\Gamma(\rho_1) + \log_2 \Tr|\rho_2^\Gamma|.
\]
In particular,
\[
D_\Gamma(\rho\otimes \Phi(d)) = D_\Gamma(\rho)+\log_2(d),
\label{eq:asymptnocat}
\]
and for any p.p.t. state $\rho'$,
\[
D_\Gamma(\rho\otimes \rho') = D_\Gamma(\rho).
\]
\end{cor}

\begin{rems}
Subtracting $D_\Gamma(\rho_1)$ from the inequality, we obtain the bound
$D_\Gamma(\rho)\le \log_2\Tr|\rho^\Gamma|$ of \cite{Horodecki1999}.  (But
see Theorem \ref{thm:newbound} below.)  See also \cite{VidalWerner}, for an
independent rederivation.
\end{rems}

\begin{rems}
For other classes of operations, \eqref{eq:asymptnocat} is known only when
$D(\rho)>0$. \cite{Bennettetal}
\end{rems}

Since Definition \ref{def:FGamma1} maximizes over all p.p.t. operations, we
can obtain relations between different values of $\rho$ and (integral) $K$
by composing with appropriate p.p.t. operations.  The next two results
extend this.  We recall from \cite{Rains:U} that for a superoperator
$\Psi$, $\Psi^\Gamma$ is defined by
\[
\Psi^\Gamma(\rho) = \Psi(\rho^\Gamma)^\Gamma.
\]
Note that $\Omega(\Psi^\Gamma) = \Omega(\Psi)^{W\otimes W'}$, and thus
$\Psi$ is p.p.t. if and only if $\Psi$ and $\Psi^\Gamma$ are completely
positive.

\begin{thm}\label{thm:psimonotone}
For any state $\rho$, any $K>0$, and any trace-preserving superoperator
$\Psi$ such that both $\Psi$ and $\Psi^\Gamma$ are positive,
\[
F_\Gamma(\Psi(\rho);K) \le F_\Gamma(\rho;K).
\]
\end{thm}

\begin{proof} (First proof)
Let $F$ be primal optimal for $F(\Psi(\rho);K)$.  Then
$\Psi^*(F)$ is primal feasible for $F(\rho;K)$, so
\[
F_\Gamma(\Psi(\rho);K) = \Tr(F \Psi(\rho)) = \Tr(\Psi^*(F) \rho)
\le F_\Gamma(\rho;K).
\]

(Second proof) Let $D$ be dual optimal for $F_\Gamma(\rho;K)$.
Then
\begin{align}
F_\Gamma(\Psi(\rho);K)
&\le
\Tr(\Psi(\rho)-\Psi(D))_+ + 1/K \Tr |\Psi(D)^\Gamma|\\
&\le
\Tr \Psi((\rho-D)_+) + 1/K \Tr \Psi^\Gamma(|D^\Gamma|)\\
&=
\Tr(\rho-D)_+ + 1/K \Tr|D^\Gamma|\\
&=F_\Gamma(\rho;K).
\end{align}
Here we used the facts that for a positive superoperator $\Psi$ and
an arbitrary operator $\rho$,
\[
\Tr(\Psi(\rho)_+)\le \Tr(\Psi(\rho_+))\text{\quad and\quad}
\Tr(|\Psi(\rho)|)\le \Tr(\Psi(|\rho|)).
\]
\end{proof}

\begin{rem}
It follows from this that we cannot improve on the p.p.t. fidelity by using
trace-preserving superoperators $\Psi$ such that both $\Psi$ and
$\Psi^\Gamma$ are positive.  In fact, one can show using the techniques of
Section \ref{sec:symmetry} that any such operator that produces isotropic
output must in fact be p.p.t.
\end{rem}

\begin{lem}
For any state $\rho$, the function
$K F_\Gamma(\rho;K)$ is nondecreasing in $K$, while the function
\[
(K F_\Gamma(\rho;K)-1)/(K-1)
\]
is nonincreasing in $K$.
\end{lem}

\begin{proof}
We first consider $K F_\Gamma(\rho;K)$.
Writing $F'=KF$, we have
\[
K F_\Gamma(\rho;K) = \max_{F'} \Tr(F'\rho),
\]
with $F'$ subject to the constraints
\[
0\le F'\le K,\quad -1\le F^{\prime\Gamma}\le 1.
\]
Since increasing $K$ increases the feasible set, the maximum cannot
decrease.  Dually,
\[
K F_\Gamma(\rho;K) = \min_D 
K \Tr (\rho-D)_+ + \Tr |D^\Gamma|,
\]
which is nondecreasing in $K$ for any choice of $D$.

For $(K F_\Gamma(\rho;K)-1)/(K-1)$, we proceed similarly;
taking $F'=(K F-1)/(K-1)$, we have:
\[
(K F_\Gamma(\rho;K)-1)/(K-1)
=
\max_{F'} \Tr(F'\rho)
\]
with $F'$ subject to the constraints
\[
-1/(K-1)\le F'\le 1,\quad
-2/(K-1)\le F^{\prime\Gamma}\le 0.
\]
These constraints become harder to satisfy as $K$ increases, and thus
the maximum cannot increase.  Dually,
\[
(K F_\Gamma(\rho;K)-1)/(K-1)
=
\min_D
(K\Tr (\rho-D)_+ + \Tr |D^\Gamma| - 1)/(K-1).
\]
But
\begin{align}
\frac{1}{K-1} (K\Tr (\rho-D)_+ + \Tr |D^\Gamma| - 1)
&=
\frac{1}{K-1} (K\Tr (\rho-D)_+ + \Tr(D^\Gamma) + 2 \Tr (D^\Gamma)_- - 1)\\
&=
\frac{1}{K-1} (K\Tr (\rho-D)_+ - \Tr(\rho-D) + 2 \Tr (D^\Gamma)_-)\\
&=
\Tr(\rho-D)_+ +
\frac{1}{K-1} (\Tr(\rho-D)_- + 2 \Tr (D^\Gamma)_-).
\end{align}
This, of course, is nonincreasing in $K$, so we are done.
\end{proof}

For integer $K$, this corresponds to composition by the following p.p.t.
operations:

\begin{lem}
Let $I_d(f)$ denote the isotropic state 
\[
I_d(f) := f\Phi(d) + \frac{1-f}{d^2-1} (1-\Phi(d))
\]
of dimension $d$ and fidelity $f$.
If $f\le 1/d$, then for all $K>0$,
\[
F_\Gamma(I_d(f);K) = 1/K.
\]
Otherwise, for $0<K\le d$,
\[
F_\Gamma(I_d(f);K) = 1/K + \frac{fd-1}{d-1} (1-1/K),
\]
and for $K\ge d$,
\[
F_\Gamma(I_d(f);K) = \frac{fd}{K}.
\]
\end{lem}

\begin{proof}
For the first claim, take $F = 1/K$, $D = I_d(f)$, at which point
$D^\Gamma\ge 0$, so $\Tr |D^\Gamma| = 1$.
For the second claim, take
\begin{align}
F &= 1/K + \frac{ d \Phi(d)-1}{d-1} (1-1/K)\\
D &= \frac{1-f}{d^2-1} (1+d\Phi(d)).
\end{align}
Finally, for the third claim, take
\begin{align}
F &= \Phi(d)\frac{d}{K}\\
D &= I_d(f).
\end{align}
In each case, the lower bound coming from $F$ agrees with the upper bound
coming from $D$, and thus both $F$ and $D$ are optimal.
\end{proof}

\begin{rem}
In particular, we have $F_\Gamma(I_d(f);d) = \max(1/d,f)$; the fidelity of
an entangled isotropic state cannot be increased by p.p.t. operations.
\end{rem}

It is instructive to translate the relative entropy bounds of
\cite{VedralPlenio}, \cite{Rains:U} in terms of the dual SDP.
We recall the definition
\[
S(\rho||\sigma) = -\Tr(\rho(\log_2\rho-\log_2\sigma)),
\]
and the following result:

\begin{lem}\cite[Theorem 2.2]{HiaiPetz}\label{lem:HP2.2}
Let $\rho,\sigma\in {\cal P}(V)$, with $\rho$ a state.  For
$0<\epsilon<1$ and $n\in \Z^+$, define
\[
\gamma_n(\epsilon) := -\frac{1}{n} \log_2 \min_P \Tr(\sigma^{\otimes n} P),
\]
where $P$ ranges over projection operators on $V^{\otimes n}$ such that
$\Tr(P \rho)\ge 1-\epsilon$.
Then
\begin{align}
\liminf_{n\to\infty} \gamma_n(\epsilon) &\ge S(\rho||\sigma)\\
\limsup_{n\to\infty} \gamma_n(\epsilon) &\le \frac{1}{1-\epsilon}
S(\rho||\sigma).
\end{align}
\end{lem}

\begin{rem}
In \cite{HiaiPetz}, this is stated only when $\sigma$ is a state; scale
invariance gives the result in general.  Also, if both $\rho$ and $\sigma$
are diagonal, we may restrict $P$ to be diagonal as well; this is just the
analogous result of classical information theory.
\end{rem}

We then have:

\begin{thm} \cite{Rains:U}\label{thm:oldbound}
For any state $\rho$ and any p.p.t. state $\sigma$,
\[
D_\Gamma(\rho) \le S(\rho||\sigma).
\]
\end{thm}

\begin{proof}
We need to show that for any $x>S(\rho||\sigma)$,
\[
\limsup_{n\to\infty} F_\Gamma(\rho^{\otimes n};2^{xn}) < 1.
\]
Choose $y$ between $x$ and $S(\rho||\sigma)$, and consider
the dual SDP bound with
\[
D = 2^{yn} \sigma^{\otimes n}.
\]
Then $D$ is p.p.t., so $1/K \Tr|D^\Gamma|=2^{(y-x)n} \to 0$;
the first term is bounded below 1 by the following lemma.
\end{proof}

\begin{lem}
Let $\rho$ and $\sigma$ be arbitrary states, and let $y$ be a nonnegative
real number.  Then
\[
\limsup_{n\to\infty} \Tr(\rho^{\otimes n}-2^{yn} \sigma^{\otimes n})_+
<1
\]
whenever $y>S(\rho||\sigma)$.
\end{lem}

\begin{proof} 
Let $P_n(y)$ be the projection onto the positive part of
\[
\rho^{\otimes n}-2^{yn} \sigma^{\otimes n};
\]
then we need to show that
\[
F_n(y):=
\Tr((\rho^{\otimes n}-2^{yn}\sigma^{\otimes n})P_n(y))
=
\Tr(\rho^{\otimes n}P_n(y))
-2^{yn}\Tr(\sigma^{\otimes n}P_n(y))
\]
is bounded below 1.  Fix $\epsilon$, and consider the statement
$F_n(y)\ge 1-\epsilon$.  For this to be true, we must certainly have
\begin{align}
\Tr(\rho^{\otimes n}P_n(y))&\ge 1-\epsilon\\
\Tr(\sigma^{\otimes n}P_n(y))&\le 2^{-yn}\epsilon.
\end{align}
Letting $y(\epsilon)$ be the largest value of $y$ such that these
inequalities simultaneously hold for infinitely many $n$, we conclude by
Lemma \ref{lem:HP2.2} that
\[
y(\epsilon) \le \frac{1}{1-\epsilon} S(\rho||\sigma).
\]
In particular, if $y>S(\rho||\sigma)$, then there exists $\epsilon$
such that $y>y(\epsilon)$, so
\[
\limsup_{n\to\infty} F_n(y) < 1-\epsilon
\]
as required.
\end{proof}

\begin{rem}
Similarly, using the fact that $P_n(y)$ is optimal among projections, we
can conclude from the other half of Lemma \ref{lem:HP2.2} that
$\lim_{n\to\infty} F_n(y) = 1$ when $y<S(\rho||\sigma)$.  We also
have the natural conjecture that the lemma can be strengthened to say
$\lim_{n\to\infty} F_n(y) = 0$ when $y>S(\rho||\sigma)$.
\end{rem}

This, of course, suggests that we should remove the requirement that
$\sigma$ be p.p.t.; the same proof then gives:

\begin{thm}\label{thm:newbound}
For any states $\rho$ and $\sigma$,
\[
D_\Gamma(\rho) \le S(\rho||\sigma) + \log_2 \Tr |\sigma^\Gamma|.
\]
\end{thm}

When $\sigma$ is p.p.t., we recover the previous bound, while when
$\sigma=\rho$, we obtain the bound of \cite{Horodecki1999} (see the remark
following Corollary \ref{cor:tpbounds} above).
Note that we could also have obtained this result using Theorem 1 of
\cite{Rains:U}, based on the fidelity bound of Corollary
\ref{cor:negativitybound}; this is essentially just the dual of the
above proof.\footnote{M., P., and R. Horodecki (personal communication)
have pointed out a third proof via Theorem 2 of \cite{quant-ph/9908065}; it
is reasonably straightforward to show that the new bound satisfies their
criteria for an upper bound to distillable entanglement.}  The proof given
above was chosen to emphasize the fact that any bound on distillable
entanglement can in principle be deduced from the dual SDP bound.

If we define
\[
{\cal B}(\rho,\sigma) := S(\rho||\sigma) + \log_2 \Tr|\sigma^\Gamma|,
\]
then we have:

\begin{thm}\label{thm:dboundxform}
For any states $\rho$ and $\sigma$, and any trace-preserving superoperator
$\Psi$ with both $\Psi$ and $\Psi^\Gamma$ positive,
\[
{\cal B}(\Psi(\rho),\Psi(\sigma)) \le {\cal B}(\rho,\sigma).
\]
For any other state $\rho'$ and real number $0<p<1$,
\[
{\cal B}(p \rho+(1-p)\rho',\sigma)
\le
p {\cal B}(\rho,\sigma)+
(1-p) {\cal B}(\rho',\sigma).
\]
Finally, we have in general
\[
{\cal B}(\rho\otimes \rho',\sigma\otimes \sigma') =
{\cal B}(\rho,\sigma)
+{\cal B}(\rho',\sigma').
\]
\end{thm}

\begin{proof}
Indeed, this is true for each of the functions $S(\rho||\sigma)$ and
$\log_2 \Tr|\sigma^\Gamma|$ individually, so must be true for their sum.
\end{proof}

In general, ${\cal B}$ is {\it not} convex in $\sigma$.  In
particular, we cannot assume that a local maximum of ${\cal B}$ is
necessarily a global maximum.  This is likely to make it very difficult
to explicitly compute $\min_\sigma({\cal B}(\rho,\sigma))$, although
one can still, of course, obtain bounds from any given value of $\sigma$.

%
%
%
%
%

\section{Exploiting symmetries}\label{sec:symmetry}

If the state $\rho$ has a large group of local symmetries, we can greatly
simplify the primal and dual SDPs, in several cases to the point of being
{\it linear} programs.  The key observation is that, by the proof
of Theorem \ref{thm:psimonotone}, we have:

\begin{thm}
Let $\Psi$ be a trace-preserving superoperator with both $\Psi$ and
$\Psi^\Gamma$ positive.  Then for any state $\rho=\Psi(\rho)$ and any
$K>0$, if $F$ is primal optimal and $D$ dual optimal for
$F_\Gamma(\rho;K)$, then so are $\Psi^*(F)$ and $\Psi(D)$.  In particular,
if $\Psi^2=\Psi$, we may assume that $F$ is $\Psi^*$-invariant and $D$ is
$\Psi$-invariant.
\end{thm}

\begin{cor}\label{cor:group}
Let $G$ be any closed subgroup of $U(k)\otimes U(l)$, and let $\rho$ be
a $G$-invariant state; that is, a state such that for all $U\in G$,
\[
U \rho U^\dagger = \rho.
\]
Then for any $K>0$, there exists primal optimal $F$ and dual optimal $D$
invariant under $G$.  If we further have
\[
U_0 \rho^t U_0^\dagger = \rho,
\]
for some $U_0\in U(k)\otimes U(l)$ with $U_0 \overline{G} \overline{U_0}=G$,
then we may further take
\begin{align}
U_0 F^t U_0^\dagger &= F,\\
U_0 D^t U_0^\dagger &= D.
\end{align}
\end{cor}

\begin{proof}
Let $\Psi$ be the superoperator
\[
\Psi:\rho\mapsto \int_{U\in G} U \rho U^\dagger,
\]
integrating with respect to the uniform probability measure on $G$.
This is trace-preserving, $\epsilon$-local (thus p.p.t.), and
satisfies $\Psi=\Psi^*=\Psi^2$.  The first claim thus follows from
the theorem.

Similarly, if $\Psi'$ is the superoperator
\[
\Psi':\rho \mapsto \frac{1}{2}(\Psi(\rho)+U_0 \Psi(\rho)^t U_0^\dagger),
\]
then the theorem applies to $\Psi'$.
\end{proof}

\begin{rem}
In particular, if $\rho$ is real, then we can take $U_0=1$, allowing us to
force $F$ and $D$ to be real as well.  If $\rho=U_0\rho^t U_0^\dagger$
for some $U_0$, we will say that $\rho$ is pseudo-real.
\end{rem}

To apply this, it will be helpful to work in greater generality initially.
Suppose simply that $\rho$ is a Hermitian operator invariant under a
subgroup $G\subset U(k)$; we would like an efficient representation of
$\rho$ in which it is still straightforward to test positivity.

Clearly, $\rho$ is invariant under $G$ if and only if $\rho$ commutes with
every element of $G$.  But then $\rho$ in fact commutes with the algebra
$\C[G]$ of linear combinations of elements of $G$.  In other words, $\rho$
must be an element of the centralizer algebra $A$ of $\C[G]$.
From representation theory, we have:

\begin{lem}
There exists a unitary change of basis exhibiting an isomorphism
\[
\C[G] \cong \oplus_\lambda \left(\Mat(d_\lambda,\C)\otimes 1_{m_\lambda}\right),
\]
for appropriate constants $d_\lambda$ and $m_\lambda$ such that
\begin{align}
\sum_\lambda d_\lambda^2 &= \dim(\C[G]),\\
\sum_\lambda m_\lambda d_\lambda &= k.
\end{align}
In the same basis, the centralizer algebra is given by
\[
\oplus_\lambda \left(1_{d_\lambda}\otimes \Mat(m_\lambda,\C)\right).
\]
\end{lem}

In particular, the state $\rho$ is determined by a set of Hermitian
operators $\rho_\lambda$, with dimensions $m_\lambda$; furthermore, $\rho$
is positive if and only if $\rho_\lambda$ is positive for each $\lambda$.
Pseudo-reality conditions also carry over readily: in an appropriate basis,
they produce conditions of the form (a) $\rho_\lambda$ real, (b)
$\rho_\lambda=\overline{\rho_{\lambda'}}$, or (c) $\rho_\lambda$
quaternionic.  Finally, we have the trace identity
\[
\Tr(\rho\sigma) = \sum_\lambda d_\lambda \Tr(\rho_\lambda \sigma_\lambda).
\]

In particular, our simplification of $F_\Gamma$ above can be viewed
as a special case of this, based on the following two examples:

\begin{eg}
Let $G_i$ be the subgroup of $U(d^2)$ consisting of operators $U\otimes
\overline{U}$.  Any $G_i$-invariant operator can be written in the form
\[
\rho = x \Phi(d) + y (1-\Phi(d)),
\]
with $\rho\ge 0$ iff $x,y\ge 0$.
\end{eg}

Partial-transposing the above example, we get: 

\begin{eg}
Let $G_w$ be the subgroup of $U(d^2)$ consisting of operators $U\otimes U$.
Any $G_w$-invariant operator can be written in the form
\[
\rho = \frac{x}{2} (1+d \Phi(d)^\Gamma)+\frac{y}{2}(1-d \Phi(d)^\Gamma)
\]
with $\rho\ge 0$ iff $x,y\ge 0$.
\end{eg}

Another important example is:

\begin{eg}
Let $\rho$ be a state of dimension $d$.  Then the state $\rho^{\otimes n}$
is invariant under the symmetric group $S_n$, acting by permuting the
tensor factors.  If $\rho'$ is a generic $S_n$-invariant operator, then the
blocks $\rho'_\lambda$ are in one-to-one correspondence with the degree $n$
representations of $\GL_d(\C)$, in such a way that $\rho^{\otimes n}$ maps
to the image of $\rho$ in the corresponding representation.
\end{eg}

If $\rho$ itself has symmetries, then we can simplify further.  If $A(x,y)$
is a homogeneous polynomial in two variables, then we write
\[
A(x,y)\succeq 0
\]
to denote the condition that $A$ has nonnegative coefficients; similarly,
\[
A(x,y)\preceq B(x,y)
\]
means that $B(x,y)-A(x,y)$ has nonnegative coefficients.

\begin{thm}
For any real numbers $0\le f\le 1$, $K>0$ and any integers $d>1$, $n>0$,
\[
F_\Gamma(I_d(f)^{\otimes n};K) = \max_{B,S} B(f,(1-f)/(d^2-1)),
\]
where $B(x,y)$ and $S(x,y)$ range over homogeneous polynomials of degree
$n$ such that
\begin{align}
0\preceq B(&x,y) \preceq (x+(d^2-1)y)^n\\
-\frac{1}{K} \left(\frac{d^2+d}{2} x + \frac{d^2-d}{2} y\right)^n
\preceq
S(&x,y)
\preceq
\frac{1}{K} \left(\frac{d^2+d}{2} x + \frac{d^2-d}{2} y\right)^n\\
S(x,y)
&=
B\left(\frac{(d+1)x-(d-1)y}{2},\frac{x+y}{2}\right).
\end{align}
\end{thm}

\begin{proof}
Let $F$ be primal optimal for $F_\Gamma(I_d(f)^{\otimes n};K)$ such that
$F$ is invariant under $S_n$ and $G_i^n$.  The representations of this
group are in one-to-one correspondence with the integers $0\le \lambda\le
n$, with $d_\lambda = \binom{n}{\lambda} (d^2-1)^\lambda$ and
$m_\lambda=1$.  Writing
\begin{align}
B_\lambda &= d_\lambda F_\lambda,\\
B(x,y) &= \sum_\lambda B_\lambda x^{n-\lambda} y^{\lambda},
\end{align}
we have
\begin{align}
\Tr(F I_d(f)^{\otimes n})
&=
\sum_\lambda d_\lambda F_\lambda (\rho^{\otimes n})_\lambda\\
&=
\sum_\lambda B_\lambda f^{n-\lambda} ((1-f)/(d^2-1))^{\lambda}\\
&= B(f,(1-f)/(d^2-1)).
\end{align}
We next observe that $0\le F$ iff $0\preceq B(x,y)$ and $F\le 1$ iff
\[
B(x,y) \preceq \sum_\lambda d_\lambda x^{n-\lambda} y^\lambda
=
(x+(d^2-1) y)^n.
\]

Similarly, the partial transpose $F^\Gamma$ is invariant under
$S_n$ and $G_w^n$.  Again the representations are indexed by $0\le
\lambda\le n$, with
\begin{align}
d'_\lambda &= \binom{n}{\lambda}
\left( \frac{d^2+d}{2}\right )^{n-\lambda}
\left( \frac{d^2-d}{2}\right )^{\lambda}\\
m'_\lambda &= 1
\end{align}
Defining
\begin{align}
S_\lambda &= d'_\lambda (F^\Gamma)_\lambda,\\
S(x,y) &= \sum_\lambda S_\lambda x^{n-\lambda} y^{\lambda},
\end{align}
we obtain the condition
\[
-\frac{1}{K} \left(\frac{d^2+d}{2} x + \frac{d^2-d}{2} y\right)^n
\preceq
S(x,y)
\preceq
\frac{1}{K} \left(\frac{d^2+d}{2} x + \frac{d^2-d}{2} y\right)^n.
\]

Finally, the relation between $S(x,y)$ and $B(x,y)$ obtains by noting that
\begin{align}
S(x,y)
&=
\Tr\Bigl(F^\Gamma \Bigl(\frac{x}{2} (1+d \Phi(d)^\Gamma)+\frac{y}{2}(1-d \Phi(d)^\Gamma)\Bigr)^{\otimes n}\Bigr)\\
&=
\Tr\Bigl(F \Bigl(\frac{(d+1)x-(d-1)y}{2} \Phi(d)+\frac{x+y}{2}
(1-\Phi(d))\Bigr)^{\otimes n}\Bigr)\\
&=
B\Bigl(\frac{(d+1)x-(d-1)y}{2},\frac{x+y}{2}\Bigr).
\end{align}
\end{proof}

\begin{rem}
For $d=2$, this linear program appeared in \cite{Rains:0}, as an upper
bound on the fidelity of separable distillation; the observation that it
provides a {\it lower} bound on p.p.t. distillation is new.
\end{rem}

Similarly,

\begin{thm}
Fix a real number $0\le p\le 1$ and an integer $d\ge 2$, and let $W_d(p)$
denote the Werner state
\[
W_d(p) = \frac{1-p}{d^2+d} (1+T(21)) + \frac{p}{d^2-d} (1-T(21)),
\]
where $T(21) = d \Phi(d)^\Gamma$.  Then
\[
F_\Gamma(W_d(p)^{\otimes n};K)
=
\max_{B,S}
B(\frac{2(1-p)}{d^2+d},\frac{2p}{d^2-d}),
\]
where
\begin{align}
0\preceq B(&x,y)\preceq \left(\frac{d^2+d}{2} x + \frac{d^2-d}{2} y\right)^n\\
-\frac{1}{K} (x+(d^2-1)y)^n \preceq S(&x,y)\preceq \frac{1}{K} (x+(d^2-1)y)^n\\
B(x,y)
&=
S\left(\frac{(d+1)x-(d-1)y}{2},\frac{x+y}{2}\right).
\end{align}
\end{thm}

\begin{cor}
For the antisymmetric Werner state $W_d(1)$, we have
\begin{align}
F_\Gamma(W_d(1);K) &= \min(1,\frac{d+2}{dK}).\label{eq:wernerfid}\\
D_\Gamma(W_d(1)) &= \log_2\left(\frac{d+2}{d}\right).
\end{align}
For any state $\rho$,
\begin{align}
F_\Gamma(\rho\otimes W_d(1);K)
&=
F_\Gamma(\rho;\frac{dK}{d+2}).\label{eq:werneradd}\\
D_\Gamma(\rho\otimes W_d(1))
&=
D_\Gamma(\rho) + D_\Gamma(W_d(1)).
\end{align}
\end{cor}

\begin{proof}
We observe that $\Tr|W_d(1)^\Gamma| = \frac{d+2}{d}$.  Thus
if we show that $F_\Gamma(W_d(1);\frac{d+2}{d}) = 1$, the proof
of Corollary \ref{cor:nocatalysis} will apply to give equation
\eqref{eq:werneradd}; taking $\rho=\Phi(1)$ gives equation 
\eqref{eq:wernerfid}, and the equations for $D_\Gamma$ follow immediately.
It thus remains to show $F_\Gamma(W_d(1);\frac{d+2}{d}) \ge 1$ (since the
other inequality is immediate).

Taking
\begin{align}
B(x,y)&=\left(\frac{d^2+d}{2}\right) \left(\frac{d-2}{d+2}\right) x +
 \frac{d^2-d}{2} y,\\
S(x,y)&=\frac{d}{d+2} (-x+(d^2-1)y),
\end{align}
we find
\[
F_\Gamma(W_d(1);\frac{d+2}{d})
\ge
B(0,\frac{2}{d^2-d})
=
1.
\]
\end{proof}

Similar results apply to ``iso-Werner'' states---states which are linear
combinations of $1$, $T(21)$, and $\Phi(d)$ (invariant under $O\otimes O$
with $O\in O(d)$)---and Bell-diagonal states---states on $\C^{2\times 2}$
which are linear combinations of $\Phi(2)$ and $\sigma_w \Phi(2)
\sigma_w^{-1}$ for $w\in \{x,y,z\}$.

Using Theorem \ref{thm:dboundxform}, we can apply the argument of Corollary
\ref{cor:group} to conclude that when minimizing ${\cal B}(\rho,\sigma)$,
we may insist that $\sigma$ possess the symmetries of $\rho$.  When $\rho$
is isotropic, we learn nothing new (the earlier bound (\cite{Rains:0},
\cite{VedralPlenio}, \cite{Rains:U}) is unchanged), but when $\rho$ is
Werner, we obtain:

\begin{cor}
Fix a real number $0\le p\le 1$ and an integer $d>2$.  Then
\[
D_\Gamma(W_d(p))
\le
\min_\sigma {\cal B}(W_d(p),\sigma)
=
\begin{cases}
0& 0\le p\le \frac{1}{2}\\
1 + p\log_2(p) + (1-p)\log_2(1-p)& \frac{1}{2}\le p\le \frac{1}{2}+\frac{1}{d}\\
\log_2 \left(\frac{d-2}{d}\right) + p \log_2 \left(\frac{d+2}{d-2}\right)
& \frac{1}{2}+\frac{1}{d}\le p\le 1.
\end{cases}
\]
\end{cor}

\begin{proof}
By the above argument, we may assume $\sigma = W_d(p')$.  Now,
\[
{\cal B}(W_d(p),W_d(p'))
=
p \log_2 \left( \frac{p}{p'}\right)
+
(1-p) \log_2 \left( \frac{1-p}{1-p'} \right)
+
\begin{cases}
1 & p'\le \frac{1}{2}\\
1 + \frac{2 (2p-1)}{d} & \frac{1}{2}\le p'\le 1.
\end{cases}
\]
We find that the optimal $p'$ satisfies
\[
p' =
\begin{cases}
p & 0\le p\le \frac{1}{2}\\
\frac{1}{2} & \frac{1}{2}\le p\le \frac{1}{2}+\frac{1}{d}\\
\frac{p(d-2)}{d+2-4p}&\frac{1}{2}+\frac{1}{d}\le p\le 1.
\end{cases}
\]
Plugging in, we obtain the stated bound.
\end{proof}

\begin{rems}
We observe that this bound is differentiable and convex for $0<p<1$,
and tight for $p=1$.
\end{rems}

\begin{rems}
The above bound has recently been independently derived in
\cite{quant-ph/0103096}, as the regularized relative entropy of
entanglement; that is,
\[
\lim_{n\to\infty}
\frac{1}{n}\min_{\sigma\text{ p.p.t}} S(\rho^{\otimes n}||\sigma).
\]
This suggests that the bounds of Theorems \ref{thm:oldbound}
and \ref{thm:newbound} may regularize to the same bound.
\end{rems}

\section{Hashing analogues}

One of the few known lower bounds on distillable entanglement is based
on the ``hashing'' protocol \cite{Bennettetal}; it will be instructive
to consider this bound (for p.p.t. distillation) via the present techniques.
The key point of the hashing bound is that on ``low weight'' states, it
gives fidelity close to 1, while on ``high weight'' states, it gives
fidelity close to 0.  This suggests the reasoning behind the following
proof:

\begin{thm}\label{thm:stronghash}
Fix a fidelity $\frac{1}{2}\le f\le 1$ and an integer $d>1$.  Then
\[
D_\Gamma(I_d(f))
\ge
\max(\log_2 d + f\log_2 f + (1-f) \log_2 \frac{1-f}{d+1},0)
.
\]
\end{thm}

\begin{proof}
Fix an integer $n>0$, and consider the set ${\cal P}_n$ consisting of tensor
products
\[
P = \otimes_{1\le i\le n} P_i
\]
with each $P_i \in \{\Phi(d),1-\Phi(d)\}$; note, in particular, that ${\cal
P}_n$ is a set of mutually orthogonal projections.  Since
\[
|\Phi(d)^\Gamma| = \frac{1}{d}\quad \text{and}\quad
\frac{d-1}{d} \le |1-\Phi(d)^\Gamma|\le \frac{d+1}{d},
\]
we have
\[
|P^\Gamma|\le d^{-n} (d+1)^{\wgt(P)},
\]
where we define $\wgt(P)$ to be the number of factors equal to $1-\Phi(d)$.

Let us then define an operator
\[
F_n(w) = \sum_{\substack{P\in {\cal P}_n\\ \wgt(P) \le w}} P.
\]
We observe that $F_n(w)$ is a projection, so $0\le F_n(w)\le 1$, and that
\[
\Tr(F_n(w) (I_d(f)^{\otimes n})
=
\sum_{0\le i\le w} \binom{n}{i} f^{n-i} (1-f)^i,
\]
which tends to $1$ as $n\to\infty$ as long as
\[
\omega := \lim_{n\to\infty} w/n > 1-f.
\]
We also compute
\[
|F_n(w)^\Gamma|
\le
\sum_{\substack{P\in {\cal P}_n\\ \wgt(P) \le w}} 
|P^\Gamma|
\le
d^{-n} \sum_{0\le i\le w} \binom{n}{i} (d+1)^i.
\]
If we take $\omega < \frac{d+1}{d+2}$, then we obtain the limit
\[
\lim_{n\to\infty}
\frac{1}{n}\log_2(d^{-n} \sum_{0\le i\le w} \binom{n}{i} (d+1)^i) = -\omega
\log_2\omega - (1-\omega)\log_2(1-\omega) + \omega \log_2 (d+1) -
\log_2(d).
\]

But then by Theorem \ref{thm:sdp1}, we conclude that
\[
D_\Gamma(I_d(f))
\ge
\log_2(d)
+
\omega
\log_2\left(\frac{\omega}{d+1}\right) + (1-\omega)\log_2(1-\omega)
\]
whenever $1-f < \omega < \frac{d+1}{d+2}$.  Since this is decreasing over
the range, we obtain the strongest bound by taking the limit as $\omega\to
1-f$, proving the theorem.
\end{proof}

\begin{rem}
When $d=2$, this is precisely the hashing lower bound (albeit weaker, in
that it applies only to p.p.t. distillation).  However, for $d>2$, the new
bound is strictly stronger.
\end{rem}

This gives us a general technique for proving lower bounds on
p.p.t.-distillable entanglement: approximate the given state as a linear
combination of projections with well-controlled partial transposes.
Our primary application of this will be to ``maximally correlated states''
\cite{Rains:U}.  We recall that a maximally correlated operator is one of the
form
\[
\rho = \rho_\alpha:=\sum_{1\le i,j\le k} \alpha_{ij}|ii\rangle\langle jj|,
\]
for some positive Hermitian operator $\alpha$, and similarly for a maximally
correlated state.  In \cite{Rains:U}, an upper bound was given on the
p.p.t. distillable entanglement, and the conjecture was made that this
bound was tight (even for the 1-locally distillable entanglement).  We give
a partial resolution of this conjecture:

\begin{thm}
For any maximally correlated state
\[
\rho_\alpha = \sum_{i,j} \alpha_{ij} |ii\rangle\langle jj|,
\]
the p.p.t. distillable entanglement is given by the formula
\[
D_\Gamma(\rho_\alpha) =B(\alpha):= H(\alpha_{11},\alpha_{22},\dots) - S(\alpha)
\]
\end{thm}

\begin{proof}
That this is an upper bound was shown in \cite{Rains:U}, so it suffices
to prove the lower bound.  We construct a protocol in two steps.

First, suppose $\alpha$ possesses a transitive group of symmetries;
that is, a transitive group $G$ of permutations such that
\[
\alpha_{\pi(i)\pi(j)} = \alpha_{ij},\ \forall \pi\in G,\ 1\le i,j\le k.
\]
(For instance, the operator
\[
\alpha = \begin{pmatrix} \frac{1}{3}&\frac{i}{6}&\frac{-i}{6}\\
                         \frac{-i}{6}&\frac{1}{3}&\frac{i}{6}\\
                         \frac{i}{6}&\frac{-i}{6}&\frac{1}{3}
\end{pmatrix}
\]
is symmetric under the transitive group $\Z_3$ of cyclic shifts.)
We decompose
\[
\alpha^{\otimes n} = \sum_{\lambda\in (0,1]} \lambda p_\lambda(n),
\]
where $p_\lambda(n)$ is the orthogonal projection onto the
$\lambda$-eigenspace of $\alpha^{\otimes n}$.  Then $p_\lambda(n)$ is
symmetric under the transitive group $G^n$, and thus has constant diagonal.
If we similarly decompose
\[
\rho^{\otimes n}_\alpha = \sum_{\lambda\in (0,1]} \lambda P_\lambda(n),
\]
we find
\[
P_\lambda(n) = \rho_{p_\lambda(n)}.
\]
We can thus apply the following lemma to $P_\lambda(n)$.

\begin{lem}
Let $\rho=\rho_\beta$ be a maximally correlated operator of dimension
$d\times d$ such that $\beta$ has constant diagonal.
Then
\[
|\rho^\Gamma| \le \frac{\Tr(\rho)}{d}.
\]
\end{lem}

\begin{proof}
We compute
\[
\rho^\Gamma = \sum_{i,j} \beta_{ij} |ij\rangle\langle ji|.
\]
This is a block matrix with 1- and 2-dimensional blocks; we thus
immediately compute that its eigenvalues are $\beta_{ii}$ for $1\le i\le
d$, and $\pm |\beta_{ij}|$ for $1\le i<j\le d$.  Since $\beta$ is
positive, we have:
\[
|\beta_{ij}|^2 \le \beta_{ii}\beta_{jj},
\]
and thus the largest eigenvalue of $\rho^\Gamma$ in absolute value is
$\max_{1\le i\le d} \beta_{ii}$.
We thus have
\[
\max_{1\le i\le d} \beta_{ii}
=
\frac{1}{d} \sum_{1\le i\le d} \beta_{ii}
=
\frac{\Tr(\rho)}{d}.
\]
\end{proof}

Now, write
\begin{align}
P(n,x) &= \sum_{x\le \lambda\le 1} P_\lambda(n),\\
x_\epsilon &= \inf(\{x\in (0,1]: \Tr(\rho P(n,x)) > 1-\epsilon\}).
\end{align}
Then for $\epsilon>0$, we find that since
\[
|P(n,x_\epsilon)^\Gamma| \le \frac{\Tr(P(n,x_\epsilon))}{d^n},
\]
we have
\begin{align}
D_\Gamma(\rho)
&\ge
\lim_{\epsilon\to 0}
\lim_{n\to\infty} 
-\frac{1}{n} \log_2\frac{\Tr(P(n,x_\epsilon))}{d^n}\\
&=
\log_2 d - S(\alpha).
\end{align}
Since $H(\alpha_{11},\alpha_{22}\dots ) = H(1/d,1/d,\dots) = \log_2 d$,
we have proved the theorem in the symmetric case.

To reduce the general case to the symmetric case, we adapt the distillation
protocol for pure states given in \cite{puredistill}.  Given a word $w$ in
the numbers $1\dots k$, we write $\wgt_i(w)$ for the number of times $i$
appears in $w$.  Then our first step is, given
\[
\rho_\alpha^{\otimes n} = \sum_{w,w'} \left(\prod_{1\le m\le n} \alpha_{w_m,w'_m}\right)
|ww\rangle\langle w'w'|,
\]
to measure $\wgt_i$ for $1\le i\le k$.  Then the resulting (random) state
$\rho_{\alpha'}$ is maximally correlated, and $\alpha'$ admits a transitive
action of $S_n$.  Now
\[
D_\Gamma(\rho_\alpha)
=
\frac{1}{n} D_\Gamma(\rho_\alpha^{\otimes n})
\ge
\frac{1}{n} \Exp(D(\rho_{\alpha'}))
=
\frac{1}{n} \Exp(B(\alpha')),
\]
where $\Exp(\cdot)$ is the expected value, and the inequality follows from
the fact that the measurement is local, so cannot increase the expected
distillable entanglement.  It thus suffices to show that
\[
\Exp(B(\alpha')) = n B(\alpha) + o(n).
\]
Now, the measurement has at most $n^k$ different outcomes, so gives us at
most $k\log_2 n$ bits of information.  But then
\begin{align}
\Exp(H(\alpha')) &\ge n H(\alpha) - k\log_2 n,\\
\Exp(S(\alpha')) &\le n S(\alpha),
\end{align}
so we find
\[
\Exp(B(\alpha'))\ge n B(\alpha) - k\log_2 n = n B(\alpha) + o(n),
\]
as required.
\end{proof}

We also have the following general result.

\begin{thm}
Fix a finite-dimensional Hilbert space $V$, and let
\[
1_{V\otimes V} = \sum_{1\le i\le m} P_i
\]
be a partition of the identity with the $P_i$ orthogonal projections.  For
each $1\le i\le m$, let $s_i$ be the largest eigenvalue of $|P_i^\Gamma|$.
Then for any state $\rho\in {\cal P}(V\otimes V)$, we have
\[
D_\Gamma(\rho)\ge
\sum_{1\le i\le m} r_i(\log_2 r_i-\log_2 s_i),
\]
where
\[
r_i := \Tr(P_i\rho).
\]
\end{thm}

\begin{proof}
To any subset $S\subset \{1,2,\dots m\}^n$, we associate a projection
\[
P_S = \sum_{w\in S} \bigotimes_{1\le i\le n} P_{w_i},
\]
which satisfies
\[
|P_S^\Gamma| \le \sum_{w\in S} \prod_{1\le i\le n} s_{w_i}.
\]
For each $0<\epsilon<1$ and each integer $n\ge 1$, let $\beta_n(\epsilon)$
be the minimum over $S$ of the largest eigenvalue of $|P_S^\Gamma|$
subject to the constraint $\Tr(P_S \rho^{\otimes n})\ge 1-\epsilon$.
Then 
\[
D_\Gamma(\rho) \ge
\liminf_{\epsilon\to 0} \liminf_{n\to\infty} \frac{-\log_2\beta_n(\epsilon)}{n}
\]
(take $F=P_S$).  Since
\[
\Tr(P_S \rho^{\otimes n}) = \sum_{w\in S} \prod_{1\le i\le n} r_{w_i},
\]
the theorem follows by the classical analogue of Lemma \ref{lem:HP2.2}.
\end{proof}

\section{Clones}

In this section, we sketch a possible direction to take in applying the
above techniques to 1-local questions (quantum codes and distillation
protocols).

\begin{defn}
An operator $A$ on $(\C^k)^{\otimes n}$ is an ``$n$-clone'' if it can be
written in the form
\[
A = \sum_i A_i^{\otimes n}
\]
where each $A_i$ is a positive operator, or can be written as a limit
of such operators.
\end{defn}

For a permutation $\pi\in S_n$, $T(\pi)$ is the operator on
$(\C^k)^{\otimes n}$ that permutes the tensor factors by $\pi$;
when $\pi = 21\in S_2$, this agrees with our earlier notation.

\begin{thm}\label{thm:clonep}
Let $A$ be an $n$-clone.  Then for all involutions $\pi\in S_n$, and all
sets $S\subset \{1,2,\dots n\}$ that intersect each 2-cycle of $\pi$
exactly once, the following operator is positive:
\[
(A T(\pi))^{\Gamma_S}.
\]
\end{thm}

\begin{proof}
Since nonnegative linear combinations and limits of positive operators are
positive, it suffices to prove the result for $A = A_0^{\otimes n}$.  In
that case, $(A T(\pi))^{\Gamma_S}$ factors as a tensor product of the
following operators:
\[
A,\ A^t,\ \text{and}\ ((A\otimes A) T(21))^{\Gamma_2}.
\]
The first two are clearly positive; that the third is positive
is a special case of the following lemma.
\end{proof}

\begin{lem}
For any operator $A$ (not necessarily Hermitian), the operator
\[
((A\otimes A^\dagger) T(21))^{\Gamma_2}
\]
is positive.
\end{lem}

\begin{proof}
We have
\[
((A\otimes A^\dagger) T(21))^{\Gamma_2}
=
((A\otimes 1) T(21) (A^\dagger\otimes 1))^{\Gamma_2}
=
(A\otimes 1) (T(21))^{\Gamma_2}  (A^\dagger\otimes 1).
\]
Since $T(21)^{\Gamma_2} = \dim(A) \Phi(\dim(A)) \ge 0$, the result
follows.
\end{proof}

For instance, let $C$ be a quantum code of length $n$ over an alphabet of
size $k$, and consider the following average over codes equivalent to $C$:
\[
W(C) = \Exp_{C'\sim C} (P_{C'}\otimes P_{C'}).
\]
This is clearly a 2-clone, so we conclude that the following operators
are positive:
\[
W(C),\ W(C)^\Gamma,\ (W(C) T(21))^\Gamma.
\]
We also find that $W(C)$ is invariant under operators of the form $U\otimes
U$, with $U$ in the semidirect product of $S_n$ acting on $U(k)^{\otimes
n}$.  Thus using the techniques of Section \ref{sec:symmetry}, we conclude
that the three given operators are positive if and only if the following
three polynomials have nonnegative coefficients:
\begin{align}
S_C(x,y) := \Tr(W(C) &(x \frac{1-T(21)}{2} + y \frac{1+T(21)}{2})^{\otimes n})\\
B_C(x,y) := \Tr(W(C)^\Gamma
&((x \frac{1}{d} T(21) + y (1-\frac{1}{d} T(21)))^{\otimes n})^\Gamma)\\
A_C(x,y) := \Tr((W(C) T(21))^\Gamma
&((x \frac{1}{d} T(21) + y (1-\frac{1}{d} T(21)))^{\otimes n})^\Gamma).
\end{align}
Using the fact that $\Tr(M^\Gamma N^\Gamma) = \Tr(MN)$, we find:
\begin{align}
S_C(x,y) &= A'_C(\frac{x+y}{2},\frac{y-x}{2}),\\
B_C(x,y) &= A'_C(y,\frac{x-y}{d}),\\
A_C(x,y) &= A'_C(\frac{x-y}{d},y),
\end{align}
where
\[
A'_C(x,y) := \Tr(W(C) (x + y T(21))^{\otimes n}).
\]
In other words, these are precisely the weight enumerators of $C$
(\cite{ShorLaflamme}, \cite{Rains:K}, \cite{Rains:W}).  In the full linear
programming bound for quantum codes, there is an additional inequality:
\[
B_C(x,y) - \frac{1}{\dim(C)} A_C(x,y) \ge 0.
\]
To prove this, we simply extend $C$ to a self-dual code $C^+$ by encoding
half of $\Phi(\dim(C))$ into $C$.  We then have
\[
A'_{C^+}(x,y,u,v) = A'_C(x,y)u + A'_C(y,x)v,
\]
so
\begin{align}
S_{C^+}(x,y,u,v) &= \frac{S(x,y)-S(-x,y)}{2} u +
                    \frac{S(x,y) + S(-x,y)}{2} v\\
B_{C^+}(x,y,u,v) &= \frac{1}{\dim(C)} A_C(x,y) u +
(B_C(x,y) - \frac{1}{\dim(C)} A_C(x,y)) v\\
A_{C^+}(x,y,u,v) &= \frac{1}{\dim(C)} A_C(x,y) u +
(B_C(x,y) - \frac{1}{\dim(C)} A_C(x,y)) v.
\end{align}
In particular, the polynomial $B_C(x,y)-\frac{1}{\dim(C)} A_C(x,y)$ must
have nonnegative coefficients.

We can thus extend the linear programming bound to higher-order invariants
(\cite{Rains:Y}) by using the relevant symmetry group to decompose the
operators attached to
\[
W_l(C^+) = \Exp_{C'\sim C^+} P^{\otimes l}_{C'}
\]
by Theorem \ref{thm:clonep}.  Note that
since $W_l(C^+) T(\pi) = W_l(C^+)$ for $\pi\in S_n$, we have only $\lfloor
\frac{l}{2}\rfloor+1$ operators to consider.

Another application of the clone concept is to 1-local operations.  
Fix a Hilbert space $V_A\otimes V_B$, and consider the 1-local operation
\[
\Psi = \sum_i {\cal A}_i\otimes {\cal B}_i,\label{eq:1locdec}
\]
where ${\cal B}_i$ are operations, and ${\cal A}_i$ are completely
positive superoperators such that $\sum_i {\cal A}_i$ is an operation.
Then as remarked in \cite{Bennettetal}, we can extend $\Psi$ to an
operation on the larger Hilbert space $V_A\otimes V_B^{\otimes n}$
by simply taking
\[
\Psi^{(n)} = \sum_i {\cal A}_i \otimes {\cal B}_i^{\otimes n}.
\]
Note that this depends not just on $\Psi$ but also on the specific
decomposition \eqref{eq:1locdec}.  The following is straightforward:

\begin{lem}\label{lem:clonable}
For any 1-local operation $\Psi$, any integer $n>1$, and
any vector $v\in V_A\otimes V_A$,
the operator
\[
\Tr_A((|v\rangle\langle v|\otimes 1)
(\Psi^{(n)}(\Phi(V_A\otimes V_B^{\otimes n}))))
\]
is an $n$-clone.
\end{lem}

Using Theorem \ref{thm:clonep}, we obtain a number of semidefiniteness
constraints that $\Psi^{(2)}$ must satisfy; these constraints can in
principle be used to obtain bounds on 1-local distillation.  (For instance,
the argument of \cite{Bennettetal} can be restated in these terms, although
we have not done so.)  Unfortunately, the resulting semidefinite programs
tend to be fairly complicated, and thus further ideas would seem to be
needed.  We also note that the cloning argument is quite fragile; if we
define a ``catalyzed'' fidelity
\[
\tilde{F}_1(\rho;K) = \limsup_{d\to\infty} F_1(\rho\otimes \Phi(d);dK),
\]
after Corollary \ref{cor:nocatalysis}, then we can no longer directly use
cloning to bound the corresponding distillable entanglement.

We close with the following new application of the cloning argument:

\begin{thm}
Fix a pair of integers $1<K<d$.  Then for all fidelities $1/d<f<1$, we
have the strict inequality
\[
F_1(I_d(f);K) < 1/K + \frac{fd-1}{d-1} (1-1/K).
\]
\end{thm}

\begin{proof}
Suppose we had equality.  A protocol $\Psi$ attaining this bound would
certainly have to be p.p.t.; thus, if we apply this protocol to $I_d(f')$,
the output fidelity will take the form $F(f')=a f' + b$ for some constants
$a$ and $b$, or equivalently
\[
F(f') = \frac{d-f'd}{d-1} a' + \frac{f'd-1}{d-1} b',
\]
for constants $a'$, $b'$.
Evaluating this at $f'=1/d$, $f'=1$, we find:
\[
a' \le 1/K,\quad b' \le 1.
\]
On the other hand, at $f'=f$, we have
\[
F(f) = (1/K)\frac{d-f'd}{d-1} + \frac{f'd-1}{d-1}.
\]
Since the coefficients are both positive, we conclude that $a'=1/K$,
$b'=1$.  In particular, $\Psi$ must take $I_d(1)=\Phi(d)$ to $I_K(1)=\Phi(K)$.

Now, consider the action of $\Psi^{(2)}$ on the state $\Phi(d)\otimes
\frac{1}{d} 1_d$.  Since $\Psi$ takes the pure state $\Phi(d)$ to the pure
state $\Phi(K)$, we conclude that $\Psi^{(2)}$ must take $\Phi(d)\otimes
\frac{1}{d} 1_d$ to a state of the form
$\Phi(K)\otimes X$;
by symmetry, we conclude that $X = \frac{1}{K} 1_K$.  But then tracing away
the other copy of $V_B$, we find that $\Psi$ takes $I_d(1/d^2)$ to
$I_K(1/K^2)$.  On the other hand, we have
\[
F(1/d^2) = \frac{d+1-K}{dK} > \frac{1}{K^2}.
\]
We thus obtain a contradiction, and the theorem follows.
\end{proof}

\begin{rem}
From \cite{Rains:T}, it follows that
\[
F_1(I_d(f);K) \ge 1/K^2 + \frac{fd^2-1}{d^2-1} (1-1/K^2)
\]
whenever $0<K\le d$.  Is this lower bound tight?
\end{rem}


\end{document}